\documentclass{article}
\usepackage[utf8]{inputenc}
\usepackage{authblk}
\usepackage{times}
\usepackage[breaklinks=true]{hyperref}
\usepackage{graphicx}
\usepackage{color}
\usepackage{amsmath}
\usepackage{eurosym}
\usepackage{amsfonts}

\begin{document}

\title{Discrete Element Simulations and Machine Learning for Improving
  the Performance of Dry Catalyst Continuous Impregnation Processes}

\author[1]{Joseph Shovlin}
\author[2]{Kuang Liu}
\author[1]{Yangyang Shen}
\author[1]{Bill Borghard}
\author[2]{Hern\'an A. Makse}

\author[1,3]{Maria S. Tomassone}

\affil[1]{Department of Chemical and Biochemical Engineering Rutgers,
  the State University of New Jersey 98 Brett Rd. Piscataway, NJ
  08854, USA}

\affil[2]{ Levich Institute and Physics Department, City College of
  New York, NY 10031, USA}

\affil[3]{Corresponding author: Maria S. Tomassone, Tel.: +1
848-445-2972, E-mail address: silvina@soe.rutgers.edu}


\date{Keywords: Catalyst impregnation; Discrete element method; 
  Machine learning;  Lasso;  Granular mixing and flow}

\maketitle

\begin{abstract}

In this work, discrete element method (DEM) simulations coupled with
machine learning are used to study the process of dry impregnation.
Our results show that the particle bed contains two regimes. Regime 1
shows smaller inclination angles and a larger mass hold-up which
implies more forces restricting the particle movement. Regime 2
reveals larger inclination angles and rotational speeds and a smaller
mass holdup, which indicates a smaller bed height. Using Machine
learning, we found a general function for the Relative Standard
Deviation (RSD) as a function of time, angle of inclination and speed
of rotation for both even and uneven flow rates for a full range of
the parameters fed in the LASSO algorithm.  Machine learning gives
insight on both regimes and reveals that for low RPM and low angles,
uneven spraying gives a lower RSD which is consistent with our
observations of the DEM studies.


\end{abstract}
\clearpage

\section{Introduction}

The impregnation of metal solutions onto porous catalyst supports is
one of the key steps for preparing industrial metal-supported
catalysts \cite{1,2,3}. In this process, metal salts or complexes are
firstly dissolved in an aqueous solution \cite{4,5,6}. The volume of
the metal solution applied is the same as or less than the total pore
volume of the catalyst support. When conducting dry impregnation in a
rotating vessel, the metal solution is typically sprayed over a
granular bed containing porous catalyst support such as alumina
(Al2O3) or silica (SiO2) \cite{7,8,9}. During spraying, capillary
action draws the metal solution into the pores and metal complexes are
adsorbed onto the high surface area support particles. The operation
typically takes 30 to 60 minutes which includes impregnation time and
subsequent mixing time. After impregnation, the catalyst support is
dried, calcined, and further pretreated to its desired active form. In
some cases, this process (impregnation, drying, calcination) is
repeated in order to add more metal to the support.

Many types of granular mixers are used in dry impregnation in batch
processes, including double cone blenders, V-blenders, and rotating
drum mixers \cite{1,2,3}. Each blender has its own unique advantages
and disadvantages to mixing, from dead zones to poor directional
mixing. Ideally, the granular catalyst support should be exposed to a
homogeneous distribution of fluid, resulting in adequate content
uniformity. However, as the liquid penetrates into the dry support,
the density and cohesion of the particles increase accordingly. A
problem may arise as excess liquid forms liquid bridges between
particles, which causes wet cohesive forces and disrupts the flow, and
further affects granular mixing processes \cite{6}.

Lately continuous processes have emerged as a new alternative to batch
processes
\cite{kumar,dubey,gao2012periodic,gao2012,sarkar,suzzi,shirazian,behjani}.
In a continuous process, the material is typically fed into a reactor
from a feeder box at one end, flow and tumble within the reactor and
exit at the other end where it is collected in a collection
box. Continuous processes have several advantages with respect to a
batch process, such as for example better efficiency (a continuous
impregnator can process from 75 to 1200kg/hr), reduced processing
time, increased throughput, improved robustness. There has been a lot
of attention in the literature on continuous processes, however not
much has been done on continuous impregnation. Kumar \cite {kumar} and
Dubey \cite{dubey} have studied continuous coating and blending of
particles. However, non-porous particles have been considered, and the
liquid considered was not spraying droplets.  Gao and coworkers have
studied periodic section modeling of convective continuous powder
mixing processes \cite{gao2012periodic,gao2012}. This study does not
consider impregnation of liquid. Sarkar and collaborators
\cite{sarkar} have done DEM simulations on a comparison of flow
microdynamics for a continuous mixer, and Suzzi {\it et al.}
\cite{suzzi} did simulations of continuous tablet coating.  Shirazian
     {\it et al.}  \cite{shirazian} and Behjani {\it et al.}
     \cite{behjani} have both worked on DEM simulations on continuous
     granulation.  Despite the amount of work on continuous processes
     there has not been any studies focusing on continuous
     impregnation of solutions on porous catalyst particles.  To the
     best of our knowledge, this is the first work on DEM simulations
     of continuous impregnation on porous particles.

Due to the large volumes and high cost associated with producing
heterogeneous catalysts in continuous impregnation, several open
questions regarding the optimization of this process still remain
unanswered: {\it (i)} how mixing and flow are affected when the
particles have a certain degree of moisture or are saturated with
fluid, {\it (ii)} how to improve fluid distribution to and within the
granular bed, {\it (iii)} the extent and distribution of dead zones
for a given impregnator configuration, {\it (iv)} how does the metal
transport throughout the granular bed when compared to fluid transfer,
{\it (v)} can the mixing and content uniformity be improved by using
alternate geometries, such as baffles, spray pattern, or systems, {\it
  (vi)} how does the rotation pattern, nozzle and spray pattern, as
well as the morphology of the catalyst support affect mixing profiles
within the granular bed, and finally, {\it (vii)} what are the rules
of thumb for process scale-up?

In this work, catalyst dry impregnation is explored by applying modern
computational techniques and machine learning.  The
discrete element method (DEM) has been increasingly used to study
granular materials and particle systems \cite{10,11}. In particular, a
commercial program known as EDEM™ (by DEM Solutions Inc.) has been
used as per the members’ recommendations for the ease of technology
transfer \cite{12,13,14,15,16}. The capabilities of this software include
user-defined functions and various features for simulating dry
impregnation. In a typical simulation, initially, catalyst support
particles are placed in a rotating mixer, such as double cone blenders
or rotating drums, and then water droplets are continuously created by
a nozzle. A novel algorithm has been developed to allow water droplets
to transfer their mass to the catalyst support particles when they are
in contact \cite{17}. After contact, water droplets “disappear”. The catalyst
support particles have a predetermined threshold to mimic the pore
volume of the experimental catalyst. When the amount of water absorbed
exceeds this value, the excess is tracked. The water transfer
algorithm allows the excess water on a specific particle to be
transferred to another adjacent particle, at a user-specified rate.

The DEM model together with the water transfer algorithm had been
validated by a series of geometrically and parametrically identical
experiments \cite{18,19}. In both simulations and experiments,
spherical g-alumina support particles of 4.7mm diameter and 35\% or
20\% pore volume (after recycled) were considered initially. Two
distinct fill levels (30\% and 45\%) were combined with three flow
rates (1.5, 2.5, and 5 L/hr) for the double cone geometry. The
rotation speed was fixed at 25 rpm. The nozzle was set up to spray in
a conical pattern of 1/3 across the axis of rotation. The fluid
content was monitored both experimentally and computationally over
time, and the relative standard deviation of fluid distribution was
used to analyze mixing. The initial study had determined that lower
fill levels (30\% fill) and lower spray rates (1.5 L/Hr) resulted in
the best mixing and content uniformity. The results of the
computations and experiments were also found to have an excellent
agreement.

\section{Model Details}

 The two main objectives for this study are 1) obtaining a fundamental understanding
of the continuous impregnation process for particles and powders, and
2) investigate process parameters and develop scientific relationships
to increase the efficiency of continuous impregnation (optimize). Some
key parameters for the continuous impregnation process include the
residence time distribution, the axial dispersion coefficient, the
product RSD (relative standard deviation), and the mass transfer
coefficients.

We focused on DEM simulations of the continuous impregnation process
before reaching a steady state. We have done a systematic study
covering before and after a steady state was reached focusing on the relative standard deviation (RSD)
and water content of tracers particles in the collection box. The
focus was to reveal details about the parameters that affect the
homogeneity of the particle bed, such as the mean residence time (MRT)
and the residence time distribution (RTD) of the tracer particles, the
separation of the nozzles and the flow rate distribution in the
nozzles.

We studied three different  values for the rotational speed: (1RPM, 3RPM and 5RPM), and 3 different angles
(1, 3 and 5 degrees) in two different flow rate configurations (even
and uneven) with 4 nozzles. We observe the following results.
We considered two different flow
rate settings:

Setting 1: Nozzles have uneven distribution

Q1: 40\% of the total; Q2: 30\% of the total; Q3: 20\% of the total;
Q4: 10\% of the total.

Setting 2: Nozzles have even distribution: Q1: 25\% of the total; Q2:
25\% of the total; Q3: 25\% of the total; Q4: 25\% of the total.

\begin{figure}[h]
\centering
\includegraphics[width=\textwidth]{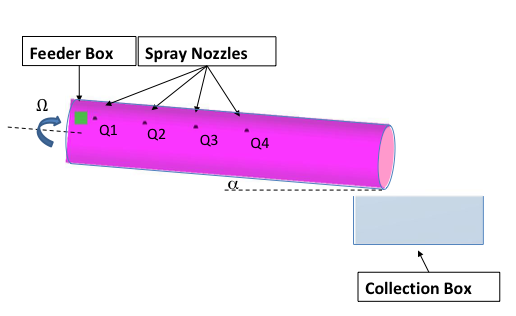}
\caption{\label{Fig:f1}\small DEM simulation setup for a continuous
  impregnation process.}
\end{figure}

Figure \ref{Fig:f1} shows the simulation setup. A full-length rotating
cylinder with a given inclination angle and RPM, and a series of
nozzles (1 to 4 nozzles) are used for the impregnation process. A
feeder box is located at one end of the cylinder and a collection box
at the other end. Particles are fed from the feeder box, flow through
the cylinder, and are collected in the collection box. The residence
time is defined as the time that particles stay inside the drum. The
probability distribution of the residence time is the Residence Time
Distribution (RTD). Figure \ref{Fig:f1} shows the depiction of the
nozzle position along the axis. In this figure, $Q_1$ to $Q_4$
indicates the flow rates through the four nozzles respectively and
$\alpha$ indicates the angle of inclination.
The nozzles were selected to be in positions that are equally spaced
at 10 cm from each other.

\section{Results of DEM simulations}

\subsection{Effect of the Inclination Angle and the Rotational Speed}

\begin{figure}[h]
\centering
\includegraphics[width=\textwidth]{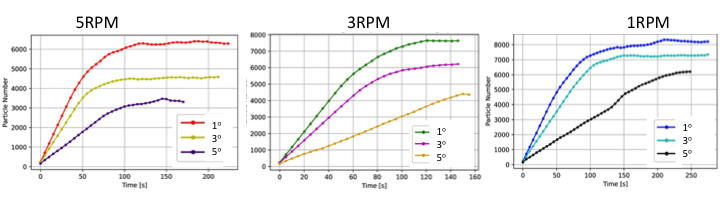}
\caption{\label{Fig:f2}\small Number of particles as a function of time for different speeds.}
\end{figure}

We first studied the number of particles (NP) in the vessel. The mass
holds up in the vessel is equal to the number of particles inside the
cylinder. The mass hold up is dependent on the flow rate and it is the
same for both patterns of flow rate. The number of particles in the
cylinder is calculated and plotted as a function of time, see Figure
\ref{Fig:f2}. The number of particles in the cylinder first increases,
and then reaches a plateau, which is considered a steady state.

We then examined the effect of inclination angle and rotational speed
and how they affect the time to reach a steady state. We varied the
inclination angle: 1, 3, and 5 degrees, and observed the time to reach
a steady state within the cylinder. The first thing that we observe is
that the number of particles inside the cylinder increases as the
rotational speed decreases. The higher the rotational speed, the
faster the steady state is reached and the shallower the bed is the
smaller the number of particles. The number of particles inside the
cylinder does not depend on the even or uneven distribution of flow
rates. Let us analyze what happens for small and large speeds.

Figure \ref{Fig:f2} shows that at 1 RPM, for 1 degree, there are
approximately 8,100 particles inside the drum, for 3 degrees there are
roughly 7,000 particles inside the drum at steady state and for 5
degrees there are roughly 6,000 particles. This plot clearly shows
that the larger the angle, the lower the number of particles that hold
up in the cylinder at the steady state (i.e. the shallower the bed)
and the sooner the steady state is reached.

At high speeds, at 5 RPM, for 1 degree, there are approximately 6,200
particles inside the drum, for 3 degrees there are roughly 4,500
particles inside the drum at the steady state and for 5 degrees there
are roughly 3,200 particles. This plot clearly shows that the larger
the angle, and the larger the speed, the lower the number of particles
that hold up in the cylinder at the steady state (i.e. the shallower
the bed) and the sooner the steady state is reached.

\subsection{Behavior After Steady State}

After reaching steady state, we focused on calculating the residence
time distribution of the particles. Residence time (RT) is defined by
the time interval between feeding and draining. For impregnation, the
time that particles stay under the spraying zone directly depends on
the RT. Tracer particles are “injected” at the inlet after steady
state is reached by labeling a specific set of particles. Equation
(\ref{et}) shows the residence time distribution, $E(t)$:

\begin{equation}
  E(t)=\frac{C(t)}{\int_{t=0}^{\infty}C(t)dt},
\label{et}
\end{equation}
where $C(t)$ is the concentration of tracer particle at the outlet as
a function of time.  While the mean residence time (MRT) can be
predicted by Sullivan’s model:
\begin{equation}
MRT = \frac{\sqrt{\phi}Lf}{\alpha
  D\omega}.
\label{mrt}
\end{equation}

\begin{figure}[h]
\centering
\includegraphics[width=\textwidth]{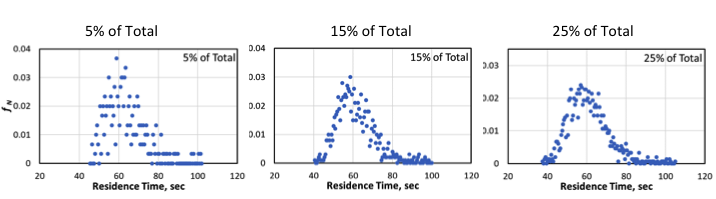}
\caption{\label{Fig:f3}\small Residence time distribution for
  different amount of tracer particles used.}
\end{figure}

In Eq. (\ref{mrt}), $\phi$ is the angle of repose, $D$ and $L$ are the
diameter and length of the cylinder, $\alpha$ is the inclination
angle, $\omega$ is the rotational speed.  The quantity of tracer was
chosen to have a good RTD. Different percentages of tracer particles
(w.r.t. Total Number of Particles) were tested. Figure \ref{Fig:f3}
shows the RTD for different tracer particle amounts. For a smaller
ratio of tracer particles, the RTD shows scattered points. For a
larger ratio of tracer particles, the RTD looks good, but it needs
more computational time. Therefore, it was found that a convenient
number for the tracer particles is 15\% of the total number of
particles. Similar ratios for the tracer particles were used by Gao
{\it et al.} \cite{gao2012}.

\begin{figure}[h]
\centering
\includegraphics[width=\textwidth]{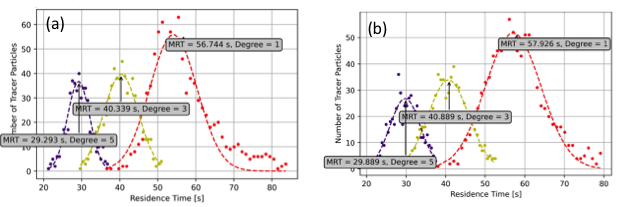}
\caption{\label{Fig:f4}\small Residence time distribution of tracer
  particles in a continuous impregnator for \textbf{5 RPM} at different
  inclinations angles, (1, 3 and 5 degrees) for (a) uneven pattern (b)
  even pattern. The points are the simulation results, and the dash
  curves are fittings with a Gaussian function.}
\end{figure}

\begin{figure}[h]
\centering
\includegraphics[width=\textwidth]{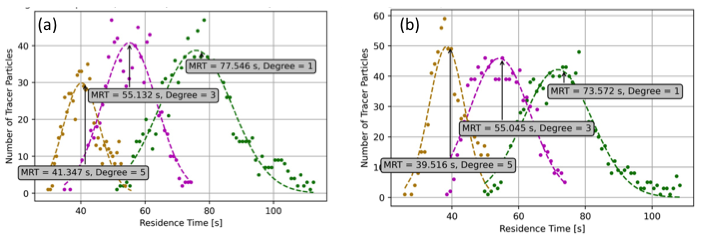}
\caption{\label{Fig:f5}\small Residence time distribution of tracer
  particles in a continuous impregnator for \textbf{3 RPM} at different
  inclinations angles, (1, 3 and 5 degrees) for (a) uneven pattern (b)
  even pattern.}
\end{figure}

\begin{figure}[h]
\centering
\includegraphics[width=\textwidth]{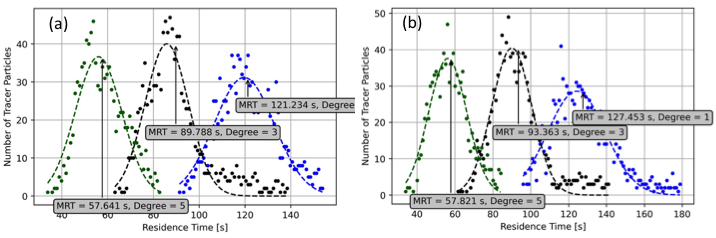}
\caption{\label{Fig:f6}\small Residence time distribution of tracer
  particles in a continuous impregnator for \textbf{1 RPM} at different
  inclinations angles, (1, 3 and 5 degrees) for (a) uneven pattern (b)
  even pattern.}
\end{figure}

We then examined the effects of rotational speed and inclination angle
on the mean residence time (MRT). Figures \ref{Fig:f4}, \ref{Fig:f5}
and \ref{Fig:f6} show the plot of the tracer particles for 5 RPM, 3 RPM,
and 1 RPM respectively as a function of the residence time, for even
and uneven patterns.

For example, looking at the plot in Figure \ref{Fig:f4} for a fixed
rotational speed (5rpm), the mean residence time (MRT) increases for
smaller angles of inclination. In other words, the smaller the angle
of inclination, the larger the MRT, which is consistent with
Sullivan’s prediction. Particles stay in the cylinder for a longer
time when the inclination angle is smaller. These findings are also
reflected for the other speeds. In other words, all the results shown
in Figures \ref{Fig:f4}, \ref{Fig:f5} and \ref{Fig:f6} show that the
larger the angle, the smaller the MRT, and the larger the speed of
rotation, the smaller the MRT. The results reflect that Sullivan’s
model can successfully predict the MRT as particles flow through the
cylinder.

We then examined the water content of the tracer particles.

It is noticeable that the larger the number of particles inside the
cylinder (i.e. the larger the mass holdup) the longer it takes to
reach 100 percent saturation. We also observe that the uneven
distributions reach 100 percent faster than the even distributions for
all cases.

\begin{figure}[h]
\centering
\includegraphics[width=\textwidth]{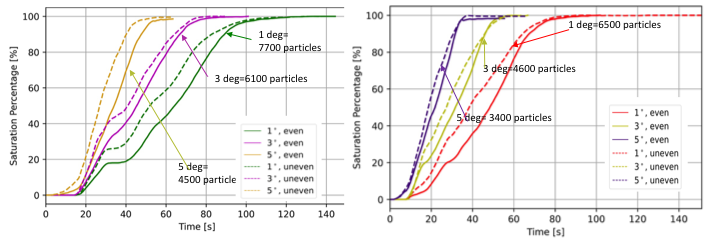}
\caption{\label{Fig:f7}\small Percentage of pore filling of the tracer
  particles as a function of time for (a) 3 RPM and (b) 5 RPM.}
\end{figure}

The larger the MRT, the longer it takes to reach uniformity. This is
not surprising, because as we see in Figure \ref{Fig:f7}a the longest
MRT corresponds to 1 degree. At this angle, the number of particles
inside the cylinder is very high ($\sim$7,700). This is one of the
highest beds possible and for that reason it takes the longest time to
fill the pore volume. Notice that the yellow curve in Figure
\ref{Fig:f7}a corresponds to $5^{\circ}$ ($\sim$4,500 particles). It
has fewer particles inside the cylinder, so it takes less time to
reach full pore volume at saturation. Notice also that the two systems
that have similar MRT (even and uneven) reach uniformity at a similar
time (solid and dash lines). Figure \ref{Fig:f7}b shows that the
larger the MRT, the better the RSD as was found previously.

\begin{figure}[h]
\centering
\includegraphics[width=\textwidth]{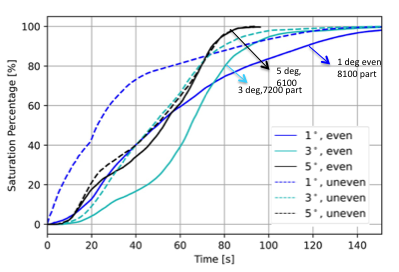}
\caption{\label{Fig:f8}\small Percentage of pore filling of the tracer
  particles for 1 RPM considered as a function of time for 1, 3, and 5
  degrees for even and uneven flow rate patterns.}
\end{figure}

We also did the case of 1 RPM. Results are shown in Figure
\ref{Fig:f8}. For 1 RPM, we also observe that the uneven distributions
reach 100 percent faster than the even distributions. However, for
1 RPM only after 60 seconds it is possible to observe the expected
trend that higher angles reach saturation faster (or vice versa:
smaller angles take more time to reach saturation).

\begin{figure}[h]
\centering
\includegraphics[width=\textwidth]{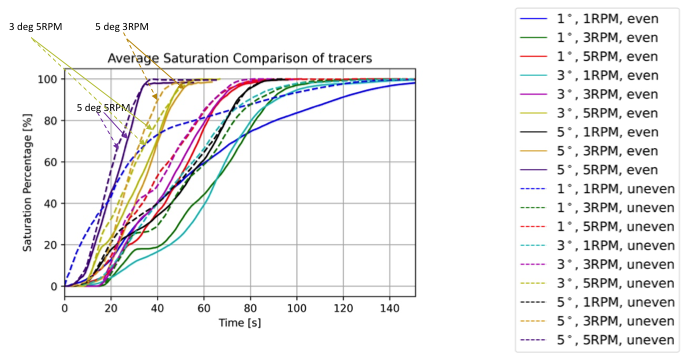}
\caption{\label{Fig:f9}\small Percentage of pore filling of the tracer
  particles for all cases considered as a function of time for 1 RPM,
  3 RPM and 5 RPM, for 1, 3 and 5 degrees for even and uneven flow rate
  patterns.}
\end{figure}

Figure \ref{Fig:f9} shows all the cases considered. We corroborate
that for higher drum speeds and higher angles the system reaches
saturation faster. As always, we can observe that the uneven pattern
configurations reach saturation faster than the even patterns of flow
rates. Higher rates and/or inclinations result in particle beds that
are so shallow that some of the sprays misses the bed.

\subsection{Relative Standard Deviation: Comparison of Even and Uneven Distribution of Flow Rates}

\begin{figure}[h]
\centering
\includegraphics[width=\textwidth]{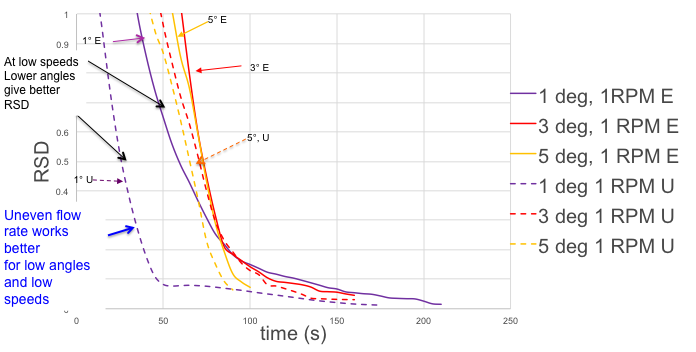}
\caption{\label{Fig:f10}\small Plot of RSD of tracers for both even and uneven for $\Omega=1$.}
\end{figure}

Figure \ref{Fig:f10} shows that for rotational speeds lower angles
give a better RSD as shown by the black arrows for both even and
uneven flow rates. We also see that at low angles of inclination (1
degree) there is a dense packing of particles at the beginning of the
vessel because the bed height is higher. So uneven flow rates favor
uniformity in this case.

\begin{figure}[h]
\centering
\includegraphics[width=\textwidth]{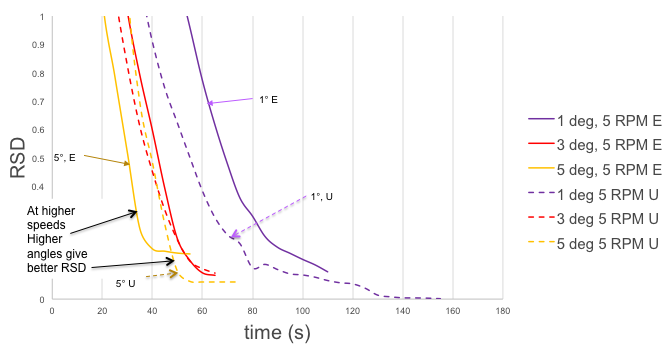}
\caption{\label{Fig:f11}\small Plot of RSD of tracers for both even and uneven for $\Omega=5$.}
\end{figure}

Figure \ref{Fig:f11} shows that at high speeds and higher angles, RSD
drops very quickly however the final values are not necessarily the
lowest as it can be observed in the solid yellow line and the dashed
yellow line. We see however that for high speeds and low angles (1
degree even and uneven), the RSD takes more time to drop (Distance
from the y axis is much longer) but the final value of the RSD is a
lot smaller as can be seen in the purple dashed and purple solid lines
because the MRT is a bit larger.

\begin{figure}[h]
\centering
\includegraphics[width=\textwidth]{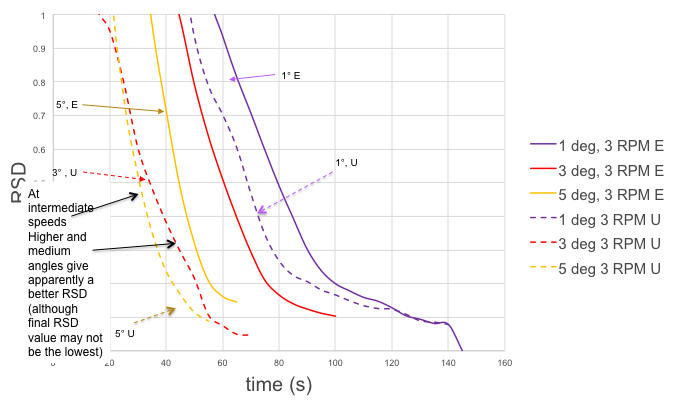}
\caption{\label{Fig:f12}\small Plot of RSD of tracers for both even and uneven for $\Omega=3$.}
\end{figure}

Figure \ref{Fig:f12} shows that at intermediate rotational speeds we
observe a somehow similar behavior to high speeds. For higher and
medium angles RSD has a quick drop but they do not give the smallest
RSD values. For low angles, medium speed such as 3 RPM does not provide
enough mixing for the large packing at the left so RSD does not drop
too quickly, however, MRT is large so ultimately RSD has the lowest
values.

\begin{figure}[h]
\centering
\includegraphics[width=\textwidth]{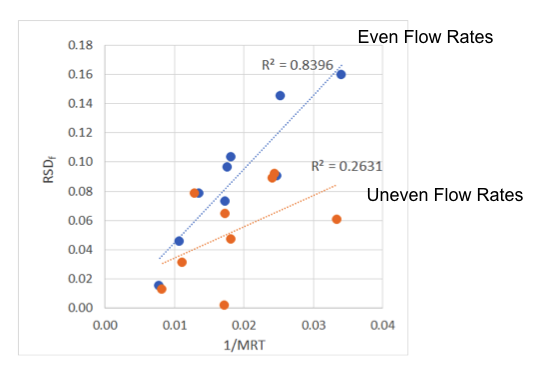}
\caption{\label{Fig:f13}\small $RSD_f$: Value of the RSD when all tracers exit the vessel(at $t_f$).}
\end{figure}

We also plotted the RSD of the tracers as a function of 1/MRT in
Figure \ref{Fig:f13}. In this figure we observe a lower (better) RSD
at higher MRT: more time, better product. This figure also shows that
the RSD values for even flow rates nicely follow the trend of
1/MRT. When we use uneven flow rates, RSD is not a monotonic function
of 1/MRT. In most cases, the RSD of tracer particles at the time the
final tracer particle exits the vessel ($RSD_f$) will decrease if MRT
increases. Increasing MRT means particles remain in the vessel
longer. Particles have more exposure to water and more time to
mix. So, increasing either RPM or angle of inclination results in
lower MRT and higher RSD (in most cases). In general, uneven flow rate
distributions improve $RSD_f$. For uneven flow rate cases,
sufficiently high RPM appears to compensate the decrease in MRT,
resulting in improved $RSD_f$.

\begin{figure}[h]
\centering
\includegraphics[width=\textwidth]{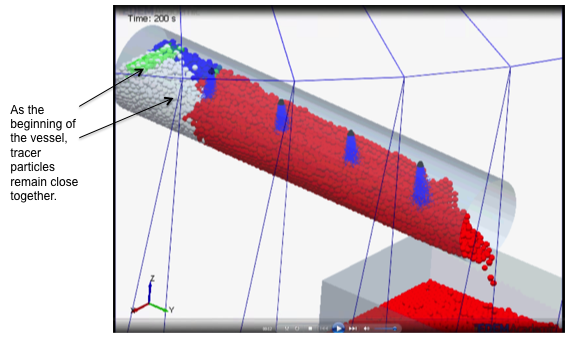}
\caption{\label{Fig:f14}\small Snapshot of 1 degree, 1 RPM uneven at
  200 seconds. At the beginning, tracer particles remain close
  together.}
\end{figure}

\begin{figure}[h]
\centering
\includegraphics[width=\textwidth]{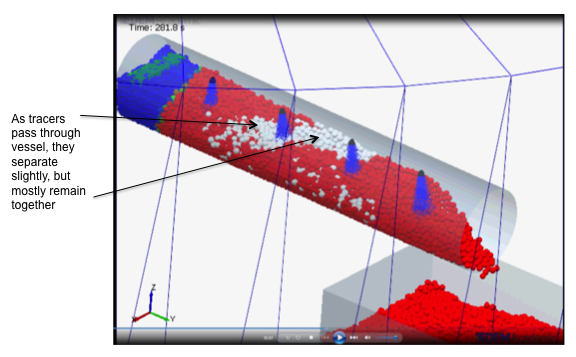}
\caption{\label{Fig:f15}\small Snapshot of 1 degree, 1 RPM, uneven at 281 seconds. We observe that as tracers pass through the vessel, they separate slightly, but mostly remain together.}
\end{figure}

\begin{figure}[h]
\centering
\includegraphics[width=\textwidth]{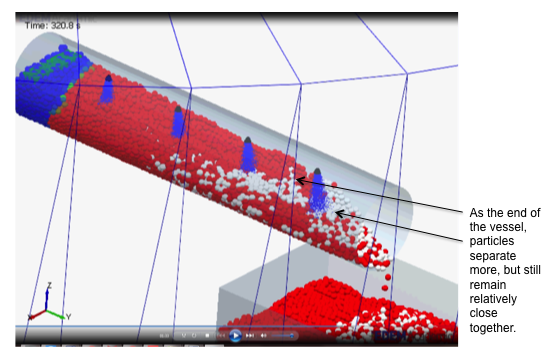}
\caption{\label{Fig:f16}\small Snapshot of 1 degree, 1 RPM, uneven at 320 seconds. At the end of the vessel, particles separate more but still remain relatively close together.}
\end{figure}

\begin{figure}[h!]
\centering
\includegraphics[width=\textwidth]{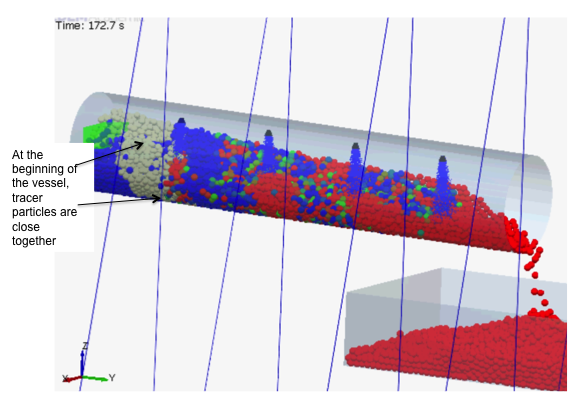}
\caption{\label{Fig:f19}\small Snapshot of 1 degree, 5 RPM, uneven at 172.7 seconds. At the beginning of the vessel, tracer particles are close together.}
\end{figure}

\begin{figure}[h!]
\centering
\includegraphics[width=\textwidth]{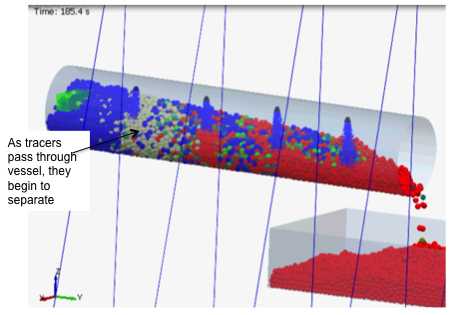}
\caption{\label{Fig:f20}\small Snapshot of 1 degree, 5 RPM uneven at 185.4 seconds. As tracers pass through the vessel, they begin to separate.}
\end{figure}

\begin{figure}[!h]
\centering
\includegraphics[width=\textwidth]{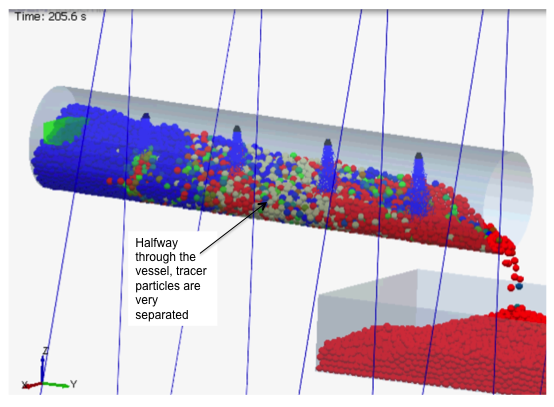}
\caption{\label{Fig:f21}\small Snapshot of 1 degree, 5 RPM, uneven at 205.6 seconds. Halfway through the vessel, tracer particles are very separated.}
\end{figure}

Figures \ref{Fig:f14} to \ref{Fig:f21} show different snapshots of the
system at different times. Figure \ref{Fig:f14} shows a snapshot of
the system at 200 seconds and low speed of the vessel (1RPM). We
observe that tracer particles remain close together at the begining of
the vessel.  Figure \ref{Fig:f15} shows a snapshot at 281 seconds
where we can see that tracers pass through the vessel, they separate
slightly but remain mostly together. A 320 seconds particles separate
a bit more but still remain relatively close together (see Figure
\ref{Fig:f16}). As we increase the speed of the vessel to 5RPM we
observe the same characteristics at the beginning of the vessel
(Figure \ref{Fig:f19}): particles remain close together and as they
pass through the vessel they separate and until they are mostly mixed
(see Figures \ref{Fig:f20} and \ref{Fig:f21}).



From Figures \ref{Fig:f14} to \ref{Fig:f21} we observed that the bed
appears to contain two regimes:

\begin{itemize}
\item Regime 1. This appears towards the beginning of the vessel. Here,
neighboring particles tend to remain close together, and the height of
the bed is large. 

\item Regime 2: As tracer particles move through vessel, they tend to
mix. Towards the end of the vessel, particles that were close together
in the beginning of the vessel are now mixed and the height of the bed
is more shallow.
\end{itemize}

\begin{figure}[!h]
\centering
\includegraphics[width=\textwidth]{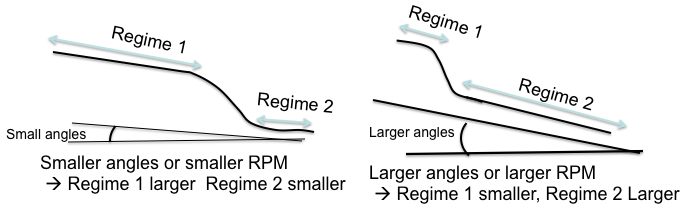}
\caption{\label{Fig:f22}\small Schematic of different regimes observed at different RPM and different angles.}
\end{figure}

Figure \ref{Fig:f22} shows a schematic of the different regimes
observed at different rotational speeds and different angles.  The two
regimes in the vessel should be treated differently. More water should
be sprayed in regime 1, where neighboring particles remain close
together. Particles will be exposed to similar amounts of water and
have more time to exchange fluid. Less water should be sprayed in
regime 2: water content uniformity will be benefited by water exchange
between particles. We obtained better results when 70\% of the total
amount of fluid is sprayed in regime 1 and 30\% in regime 2.

We also observed that the size of the regimes is a function of
inclination angle and RPM. More RPM  generates more mixing and
therefore will decrease the size of regime 1 and increase the size of
regime 2 (mixing regime). Smaller inclination angles  generate a
larger mass hold-up with stronger forces restricting the particle
movement. Larger inclination angle and/or RPM generate a smaller
mass holdup and therefore a smaller bed height, consequently a smaller
regime 1 and a larger regime 2.

\section{Machine Learning}

In order to elucidate the role of the parameters that affect the
process of continuous impregnation we use machine learning
regressions. Machine learning is a type of artificial intelligence
that allows software applications to become more accurate at
predicting outcomes without being explicitly programmed to do
so. Machine learning algorithms use data as input to predict new
output values. We focus on processing the data we got for different
values of RPM, angles of inclination, the final time for tracers, and
flow patterns, even and uneven.  There are different approaches to fit
the experimental data and build the regressed function such as LASSO
(Least Absolute Shrinkage and Selection Operator) \cite{lasso}, MARS
(Multivariate Adaptive Regression Splines) \cite{lasso}, and Neural
Network Deep Learning. Particularly the neural network usually
requires many samples which we don't have. Between LASSO and MARS, we
employed LASSO since it gives higher accuracy compared to MARS in our
case. The LASSO method is a regression analysis method that performs
both variable selection and regularization in order to enhance the
prediction, accuracy, and interpretability of the resulting
statistical model \cite{lasso}.

\subsection{Procedure using the LASSO Method}

\begin{figure}[h]
\centering
\includegraphics[width=.75\textwidth]{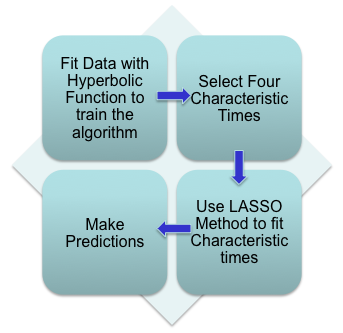}
\caption{\label{Fig:f23}\small Schematic of the method followed to use
  the LASSO method.}
\end{figure}

We follow the scheme depicted in Figure \ref{Fig:f23}, as follows.
From the simulations we obtained RSDs as functions of time for samples
with different parameters (inclination degree and rotation speed). We
first fit these data using hyperbolic functions. Then we select four
characteristic times and use the LASSO method \cite{lasso} to fit
those characteristic times from the parameters. In the end, we make
predictions with fitted functions and new parameters to get new
characteristic times, thus RSD vs. time curves of new parameters can
be determined.

\subsubsection{Data Fitting with Hyperbolic Function: Even and Uneven Spraying}

\begin{figure}[h]
\centering
\includegraphics[width=.8\textwidth]{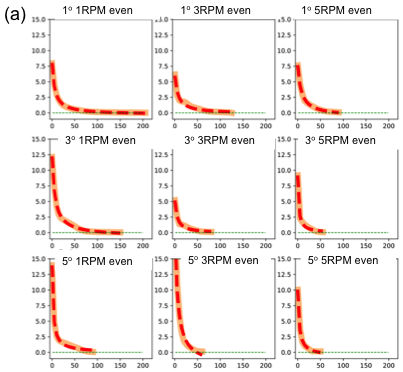}
\includegraphics[width=.8\textwidth]{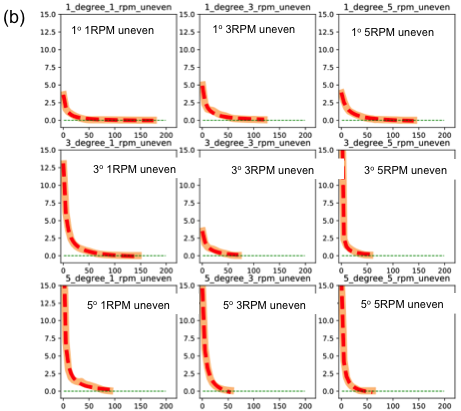}
\caption{\label{Fig:f24}\small Fitted hyperbolic functions and data
  point for (a) even and (b) uneven cases. The dashed lines are the
  fitted hyperbolic function and the orange lines are the real Data.}
\end{figure}

The first step is to get a hyperbolic function that fits the data to
have a continuous curve, since the data points are discrete and we
need a continuous curve to predict new values. Figure \ref{Fig:f24}
shows the fitted hyperbolic functions for (a) even, and (b) for uneven
spraying cases.

The hyperbolic function (red dashed line) in Figure \ref{Fig:f24} was
used to fit the RSD discrete data points obtained from our experiments
by using the following equation:
\begin{equation}
  RSD(t)=\frac{a}{t+b}+c .
  \label{rsd}
\end{equation}

\begin{figure}[h]
\centering
\includegraphics[width=\textwidth]{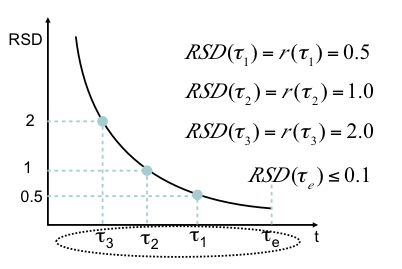}
\caption{\label{Fig:f25}\small Schematic of the Relative Standard
  Deviation as a function of time showing the different characteristic
  times $\tau_1$, $\tau_2$, $\tau_3$, and $\tau_e$. Each of these
  times are chose according to: RSD($\tau_1$)= 0.5 , RSD($\tau_2$)= 1 ,
  RSD($\tau_3$)= 2 , and RSD($\tau_e) < 0.1$ .}
\end{figure}

An hyperbola described by the equation above is determined by three
parameters $a$, $b$ and $c$. But, more generally, an hyperbola can be
determined by any three points defined on it. In order to give
parameters more realistic meanings, we extracted 3 characteristic
times $\tau_1$, $\tau_2$, $\tau_3$, and also the final time $\tau_e$
at which all the tracers exit the vessel. We will use these four
parameters to feed into the regression algorithm. Figure \ref{Fig:f25}
shows the characteristic times that are located on the time axis, and
each of these characteristic times corresponds to a pre-selected
certain value of the RSD. The goal is to find the lowest $\tau_1$,
$\tau_2$, and $\tau_3$ to obtain the highest quality product. Notice
that lower values of $\tau_e$ relates to higher production rate.

We then considered two parameters to determine the RSD function for a
simulation: the rotational speed $\Omega$ of our samples and the angle
of inclination $\alpha$ of the vessel, for both even and uneven
spraying. The goal is to find the functions $f^{1}$, $f^{2}$, $f^{3}$,
and $f^{e}$, such that $\tau_i=f^{(i)}(\Omega, \alpha)$, where
$i = 1, 2, 3, e$.

In LASSO, functions $f^{(i)}$ are parameterized by $\beta$s in the
following
way: $$\tau_i=\beta_1^i+\beta_2^i\Omega+\beta_3^i\alpha+\beta_4^i\Omega^2+\beta_5^i\alpha^2+\beta_6^i\Omega\alpha+\beta_7^i\Omega^3+\beta_8^i\Omega^2\alpha+\beta_9^i\Omega\alpha^2+\beta_10^i\alpha^3 , $$
 where $i = 1, 2, 3$. Specifically, when fitting $f^{(e)}$ we expand it
only to the second order of parameters since the fitting is already
good enough so overfitting is avoided.

In the LASSO method \cite{lasso}, the idea is to make the fit small by
making the residual squares small plus a penalty, in such a way
that:
\begin{equation}
  \min(\sum^9_{j=1}(\tau_{i,j}-\tau_{i,j}^{real})^2+\lambda\sum^{10}_{k=1}|\beta_{i,k}|) , \label{min}
\end{equation}
 where $i$ labels different characteristic time, $j$ labels all 9
samples we have, and $k$ labels 10 $\beta$s used in the fitting
functions.

\subsubsection{Predictions from Fitted Hyperbolic Functions}

Once we have fitted functions we are able to predict three
characteristic times from any given new parameters $\Omega$ and
$\alpha$. Thus we can obtain the parameters $a$, $b$, and $c$ in the
hyperbolic function from three characteristic times following:

\begin{equation}
  b=\frac{\frac{\tau_3}{\tau_3-\tau_2}-\frac{\tau_1}{2(\tau_2-\tau_1)}}{\frac{1}{2(\tau_2-\tau_1)}-\frac{1}{\tau_3-\tau_2}} , 
\end{equation}

\begin{equation}
  a=-\frac{1}{2}(\frac{(\tau_1+b)(\tau_2+b)}{2(\tau_2-\tau_1)}+\frac{(\tau_2+b)(\tau_3+b)}{\tau_3-\tau_2}) , 
  \end{equation}
    \begin{equation}
      c=\frac{1}{3}(\frac{7}{2}-\frac{a}{\tau_1+b}-\frac{a}{\tau_2+b}-\frac{a}{\tau_3+b}) .
\end{equation}

\begin{figure}[h]
\centering
\includegraphics[width=\textwidth]{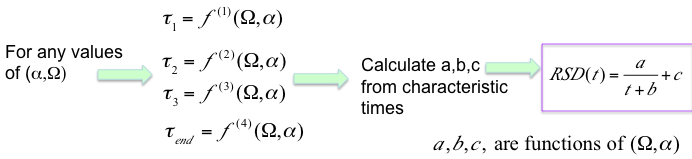}
\caption{\label{Fig:f26}\small Schematic of the process followed to
  find the RSD as a function of the characteristic time.}
\end{figure}

Using these values of $a$, $b$, and $c$ we are able to draw the full
RSD vs time curve for intermediate values of all rotational speeds and
angles within the limits of the parameters fed in the LASSO algorithm,
as shown in Figure \ref{Fig:f26}.

\section{Results: Predictions from Machine Learning}

\begin{figure}[h]
\centering
\includegraphics[width=\textwidth]{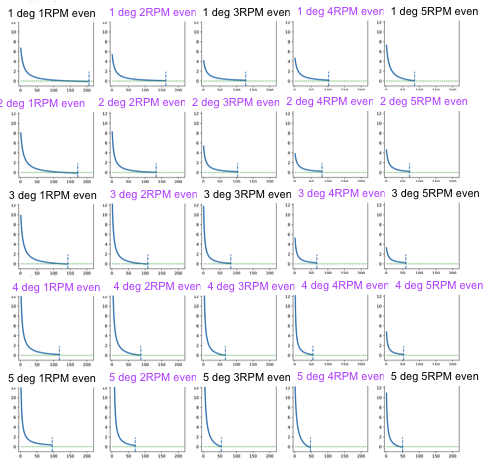}
\caption{\label{Fig:f27}\small Even cases. Panels with purple titles
  are new predictions that were not included in the data fed to the
  LASSO algorithm (i.e. these cases were never run with DEM before).}
\end{figure}

\begin{figure}[h]
\centering
\includegraphics[width=\textwidth]{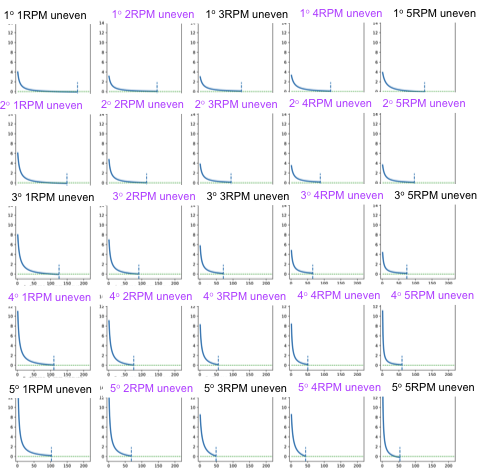}
\caption{\label{Fig:f28}\small Uneven cases. Panels with purple titles
  are new predictions never run with DEM before. We can predict
  intermediate RPMs and angles. The decay of RSD is faster for uneven
  spraying.}
\end{figure}

In Figures \ref{Fig:f27} and \ref{Fig:f28} we depict our predictions
for the even and uneven spraying cases. Using the RSD obtained as
shown in Figures \ref{Fig:f27} and \ref{Fig:f28}, we were able to make
predictions using the LASSO algorithm for the even case and uneven
cases respectively. Notice that the panels with titles in purple
correspond to the new predictions of cases that were not included in
the data fed to the algorithm (i.e. they were not run with DEM).

\begin{figure}[h]
\centering
\includegraphics[width=\textwidth]{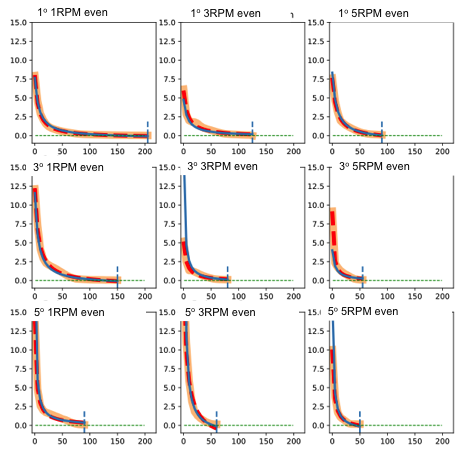}
\caption{\label{Fig:f29}\small Even cases. Red Dashed: hyperbolic fit,
  Orange: Real Data, Blue: Prediction with LASSO. We observe great
  agreement with the original data.}
\end{figure}

With machine learning we can predict intermediate cases for other RPM
values and angle values. We can observe in this figure that the decay
of the RSD is faster for uneven spraying in general.  Predictions need
to be checked for accuracy against the data and the fitted
functions. Figure \ref{Fig:f29} shows the accuracy of the predictions
compared to the experimental data and the fitted hyperbola for both
even and uneven cases respectively.

\begin{figure}[h]
\centering
\includegraphics[width=\textwidth]{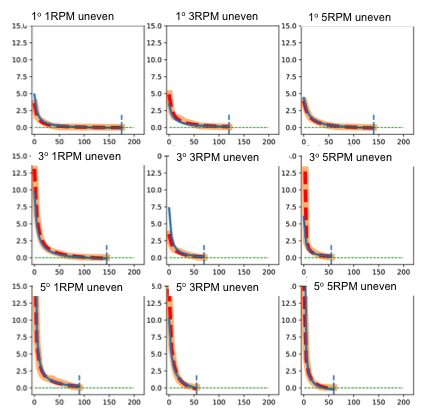}
\caption{\label{Fig:f30}\small Uneven cases. Red Dashed: hyperbolic
  fit, Orange: Real Data, Blue: Prediction with LASSO. We observe
  great agreement with the original data.}
\end{figure}

In Figure \ref{Fig:f30} we present the data for uneven cases. There is
great agreement with the original data and the fitted hyperbola. In
these figures we can observe that for low RPM and low angles, when the
bed is larger in particle density and displays a larger regime 1, the
lower RSD values correspond to the uneven cases. And for the cases of
large RPM and larger angles, when the particle bed is shallow and even
and mostly in regime 2, the lower RSD values correspond to the even
spraying scenario.

Figures \ref{Fig:f29} and \ref{Fig:f30} show that the larger the RPM
the longer it takes to achieve uniformity as seen for the cases of RPM
=1. It is interesting to notice that a slight increase in the angle
for high RPM reduces the time of uniformity by 3 as can be observed
when comparing 5 RPM and 1 degree with 5 RPM and 3 degrees.

\subsection{Qualitative Study of Regime 1 and Regime 2}

We can get a qualitative estimation of the distribution of the regimes
by looking at the particle densities in the cylinder. The particle
density $\phi$ is calculated using the feeding rate (100 particles/second),
the particle diameter $d$, the cylinder’s diameter and length $D$ and $L$,
and the MRT as:

$$\phi=\frac{NV_{part}}{V_{drum}}=\frac{100\ part/s\ MRT\ V_{part}}{V_{drum}}=\frac{100MRT\frac{4}{3}\pi(d/2)^3}{\pi(D/2)^2L}=\frac{200d^3MRT}{3D^2L}$$

\begin{figure}[h]
\centering
\includegraphics[width=\textwidth]{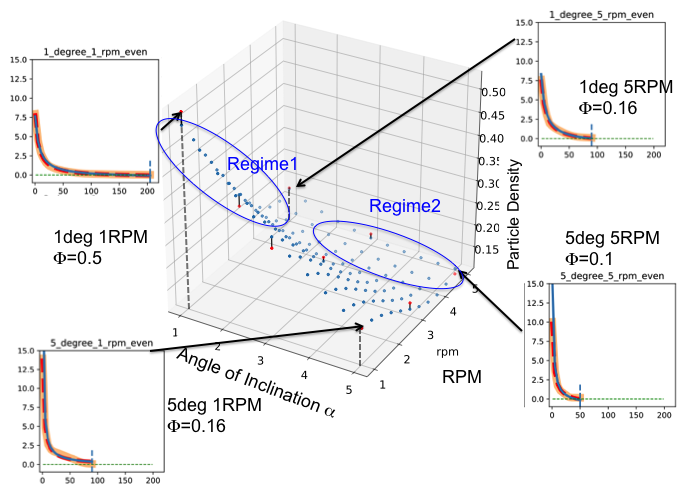}
\caption{\label{Fig:f31}\small Even case. Regimes do not depend on the spray pattern. Regimes depend on the MRT.}
\end{figure}

\begin{figure}[h]
\centering
\includegraphics[width=\textwidth]{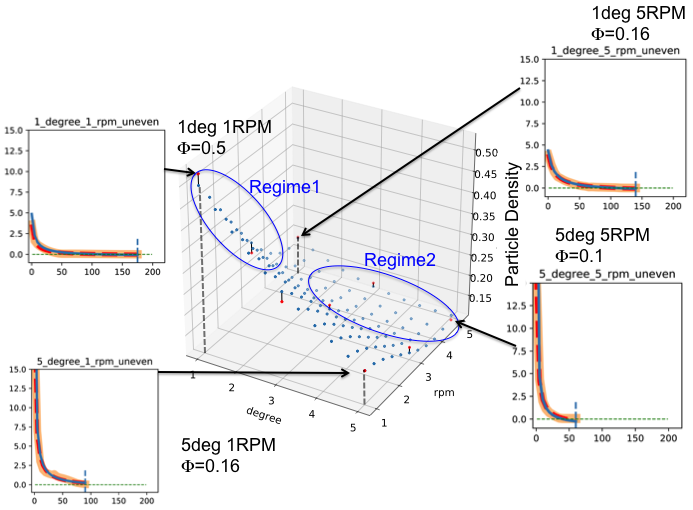}
\caption{\label{Fig:f32}\small Uneven case. Regimes do not depend on the spray pattern. Regimes depend on the MRT.}
\end{figure}

Figure \ref{Fig:f31} shows qualitatively the two different regimes, 1
and 2. Notice that for small angles of inclination and small RPM, the
particle density is the highest as indicated by the dashed line
$\phi=0.5$. We can see that the large values of the density extend
until 3 RPM and 1 degree of inclination. The RSD for these cases is not
the lowest, as indicated by the two insets. If we compare these values
with the same values in Figure \ref{Fig:f32} we see that uneven
spraying works better. That is because when there is larger density of
particles it is better to use uneven spraying, more water at the very
left of the vessel to saturate the large accumulation of particles.

Larger angles of inclination do not show a great accumulation of
particles. We can also observe in this figure that for 5 degrees and 1
RPM ($\phi=0.16$) and 5 degrees, 5 RPM, ($\phi=0.1$) the density of
particles is the lowest indicating that the system is in regime 2. We
would expect that for this regime the even distribution would give a
lower RSD compared to the uneven spraying. This can be seen in Figure
\ref{Fig:f32} in the inset 5 deg 5 RPM, where the RSD for uneven
spraying has a higher value than for the even case depicted in Figure
\ref{Fig:f31}.

\subsection{Fitting functions found by LASSO}

As mentioned earlier the goal is to predict $\tau$s that depend on the
rotational speed and the angles. Figures \ref{Fig:f33} and
\ref{Fig:f34} below shows the functions found by the LASSO algorithm
for the even cases and uneven cases.

Notice that low values of $\tau_1$, $\tau_2$ and $\tau_3$ relate to
high quality product. Also, lower $\tau_e$ relates to a higher
production rate. In Figure \ref{Fig:f34}, in general for uneven
spraying cases, we observe a faster decay of the RSD.

\subsubsection*{Functions found by LASSO: $\tau_1$}

\begin{figure}[h]
\centering
\includegraphics[width=\textwidth]{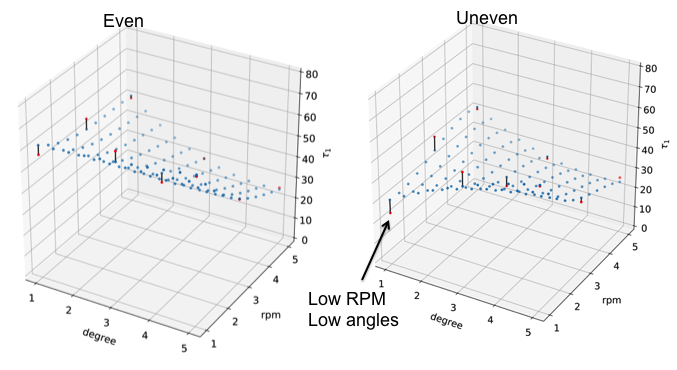}
\caption{\label{Fig:f35}\small Comparison side by side of the plots of
  $\tau_1$ as a function of RPM and angle of inclination (a) even, (b)
  uneven.}
\end{figure}

Figure \ref{Fig:f35} depicts the plots of $\tau_1$ as a function of
RPM and angle of inclination. We can observe that for uneven cases
(b), $\tau_1$ reveals that for low RPM in general gives lower
RSD. Also, this confirms that for low RPM and low angles uneven gives
better RSD as we had concluded with our DEM simulations.

Notice that, in the inset of Figure \ref{Fig:f35} we can observe that
the smallest the value of $\tau_1$, the faster RSD decays to zero,
indicating a better quality of the product.

\subsubsection*{Functions found by LASSO: $\tau_2$}

\begin{figure}[h]
\centering
\includegraphics[width=\textwidth]{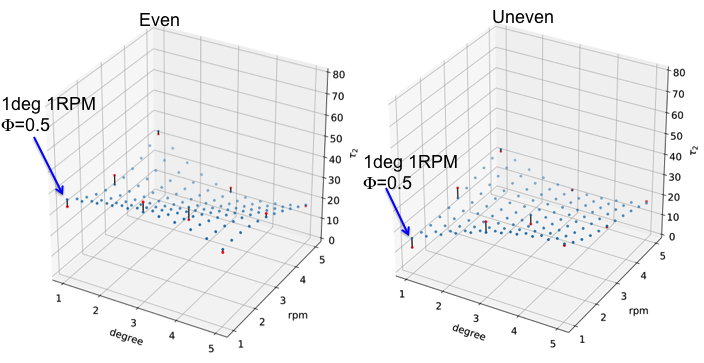}
\caption{\label{Fig:f36}\small Comparison side by side of the plots of $\tau_2$ as a function of RPM and angle of inclination (a) even, (b) uneven.}
\end{figure}

Figure \ref{Fig:f36} shows the function $\tau_2$ found with the LASSO
algorithm as a function of RPM and angle of inclination. We can
observe that for 1 degree and 1RPM, low angles and low velocities of
the drum, the uneven spraying gives a lower value of the RSD. This
makes sense because at low angles and low velocities there is a larger
Regime 1, therefore, an uneven distribution of the spraying should
give a more uniform distribution to counteract the accumulation of
particles at the left of the vessel.

\subsubsection*{Functions found by LASSO: $\tau_3$}

\begin{figure}[h]
\centering
\includegraphics[width=\textwidth]{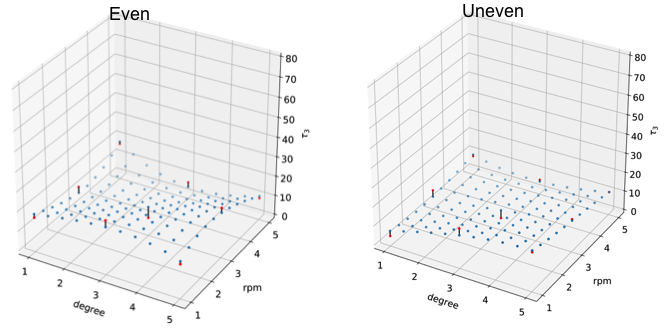}
\caption{\label{Fig:f37}\small Comparison side by side of the plots of
  $\tau_3$ as a function of RPM and angle of inclination (a) even, (b)
  uneven.}
\end{figure}

Figure \ref{Fig:f37} shows the function $\tau_3$ found with the LASSO
algorithm as a function of RPM and angle of inclination. We see that
in general for uneven spraying we observe a faster decay (quick drop)
of the RSD.

\subsubsection*{Functions found by LASSO: $\tau_e$}

\begin{figure}[h]
\centering
\includegraphics[width=\textwidth]{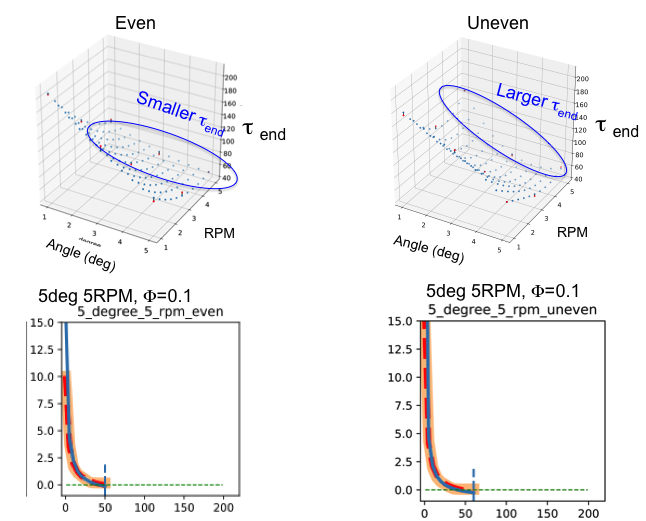}
\caption{\label{Fig:f38}\small Comparison side by side of the plots of
  $\tau_e$ as a function of RPM and angle of inclination (a) even, (b)
  uneven.}
\end{figure}

Figure \ref{Fig:f38} corresponds to $\tau_e$. In Figure
\ref{Fig:f38}(a) we see that for even spraying, the smaller values of
$\tau_e$, which indicate a really good uniformity since RSD would
decay very quickly, correspond to the highest values of the speed of
drum (high RPM) for any value of the angle of inclination. This may
indicate that when the speed of the drum is high, the height of the
bed tends to even out, and for high angles is shallow and leveled
off. Therefore, even spraying will provide more uniformity than uneven
spraying. Notice that for high RPM and high angles the $\tau_e$ is
really small in value indicating that the RSD decays very
quickly. This happens for both even and uneven although if we observe
the curves closely, the lowest value of tend for high RPM and high
angles occurs for even spraying.

\section{Conclusions}

We developed a DEM framework for a systematic study of the continuous
impregnation process. This framework allows to test and optimize the
effect of various parameters, such as the number of nozzles, the
nozzle spacing, the angle of inclination of the vessel, the wetted
area, the drum speed, the flow rate patterns, etc.

During startup, the number of particles inside the cylinder first
increases and then reaches a steady state value. The higher the
rotational speed and the higher the inclination angle, the sooner the
steady state is reached and the shallower the bed is.

In all our studies, we see that the effects of inclination angle and
rotational speed are consistent with Sullivan’s prediction: the MRT is
inversely proportional to the angle of inclination and rotational
speed. However, more studies are needed to elucidate the role of the
wetted area. The RSD for the water saturation of the tracers reveals
some general trends. We observe that at low rotational speeds lower
angles give better RSD. At high speeds higher angles give a quick drop
in RSD however they do not yield the lowest values of the RSD. At low
angles and low rotational speeds uneven flow rate helps to have a
lower RSD due to the large packing of the particles at the beginning
of the vessel.

Using machine learning we found a general function for the relative
standard deviation as a function of time, $\alpha$, $\Omega$ and for
both even and uneven flow rates. We are able to predict the full RSD
vs time for intermediate values of all rotational speeds and angles
within the limits of parameter values fed in the LASSO algorithm. In
general for uneven spraying we observe a faster decay of the RSD which
should give a better product quality. That is evident with the value
of $\tau_1$ and $\tau_2$ being lower for the uneven flow rate pattern
in general.  Machine learning also reveals that for low RPM and low
angles, since the particles are mostly in the regime 1, uneven
spraying yields much lower values of the RSD (better product quality),
which is consistent with our previous observations of the DEM
snapshots.  We see that for high RPM and high angles (particles within
the vessel are in regime 2), the particle bed is very shallow and even
works better (characteristic times are smaller).  We also verify that
regimes do not depend on the spray pattern but on the MRT, as it was
expected. The recommendation is to use uneven spraying unless the bed
is purely in Regime 2. Ultimately we developed a computational
framework that could find optimal performance within the parameter
space.

{\bf Author contributions:} KL performed the machine learning
calculations, YS and JS performed the DEM simulations, BB advised on
the project, HAM directed the machine learning part, and MST performed
some of the DEM simulations wrote the paper and directed the project.

{\bf Funding source:} This work was funded by the Rutgers Catalyst
Manufacturing Consortium (RCMC) and NSF DMR-1945909.

\clearpage

\begin{figure}[h]
\centering
\includegraphics[width=\textwidth]{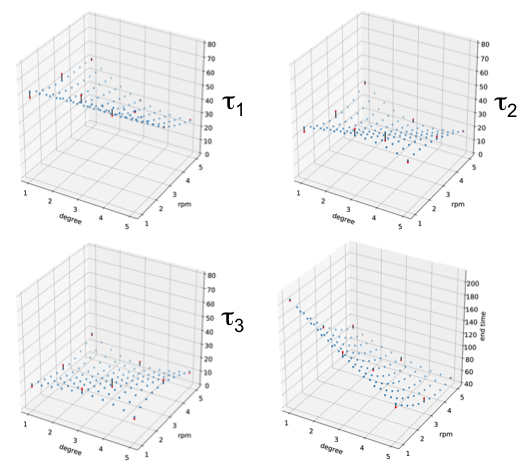}
\caption{\label{Fig:f33}\small Even cases. $\tau_1$, $\tau_2$, $\tau_3$ and $\tau_e$ as functions of $\Omega$ and $\alpha$.}
\end{figure}

\begin{figure}[h]
\centering
\includegraphics[width=\textwidth]{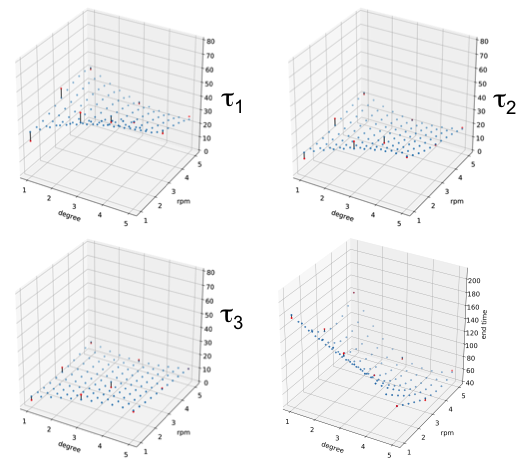}
\caption{\label{Fig:f34}\small Uneven cases. $\tau_1$, $\tau_2$, $\tau_3$ and $\tau_e$ as functions of $\Omega$ and $\alpha$.}
\end{figure}

\end{document}